\documentclass[twocolumn,conference]{IEEEtran}
\usepackage[T1]{fontenc}
\usepackage[latin9]{inputenc}
\usepackage{mathrsfs}
\usepackage{amsmath}
\usepackage{amsthm}
\usepackage{amssymb}
\usepackage{graphicx}
\usepackage[unicode=true,
 bookmarks=true,bookmarksnumbered=true,bookmarksopen=true,bookmarksopenlevel=1,
 breaklinks=false,pdfborder={0 0 0},pdfborderstyle={},backref=false,colorlinks=false]
 {hyperref}
\hypersetup{pdftitle={Your Title},
 pdfauthor={Your Name},
 pdfpagelayout=OneColumn, pdfnewwindow=true, pdfstartview=XYZ, plainpages=false}

\makeatletter
\theoremstyle{plain}
\newtheorem{thm}{\protect\theoremname}

\usepackage{cite, amsmath, amsfonts, amssymb, graphicx,lipsum}
\usepackage{amsthm}

\theoremstyle{remark}

\makeatother

\providecommand{\theoremname}{Theorem}

\begin{document}
\title{On the Secrecy Rate of Downlink NOMA in Underlay Spectrum Sharing
with Imperfect CSI}
\author{\IEEEauthorblockN{Vaibhav~Kumar\IEEEauthorrefmark{1}, Mark~F.~Flanagan\IEEEauthorrefmark{1},
Daniel~Benevides~da~Costa\IEEEauthorrefmark{2}, and Le-Nam~Tran\IEEEauthorrefmark{1}}\IEEEauthorblockA{\IEEEauthorrefmark{1}School of Electrical and Electronic Engineering,
University College Dublin, Belfield, Dublin 4, Ireland\\
\IEEEauthorrefmark{2}Future Technology Research Center, National
Yunlin University of Science and Technology, Douliu, Yunlin 64002,
Taiwan \\
Email: \{vaibhav.kumar, mark.flanagan, danielbcosta\}@ieee.org, nam.tran@ucd.ie}}
\IEEEspecialpapernotice{Invited Paper}

\maketitle
{\let\thefootnote\relax\footnotetext{This publication has emanated from research conducted with the financial support of Science Foundation Ireland (SFI) and is co-funded under the European Regional Development Fund under Grant Number 17/CDA/4786.}}
\begin{abstract}
In this paper, we present the ergodic sum secrecy rate (ESSR) analysis
of an underlay spectrum sharing non-orthogonal multiple access (NOMA)
system. We consider the scenario where the power transmitted by the
secondary transmitter (ST) is constrained by the peak tolerable interference
at multiple primary receivers (PRs) as well as the maximum transmit
power of the ST. The effect of channel estimation error is also taken
into account in our analysis. We derive exact and asymptotic closed-form
expressions for the ESSR of the downlink NOMA system, and show that
the performance can be classified into two distinct regimes, i.e.,
it is dictated either by the interference constraint or by the power
constraint. Our results confirm the superiority of the NOMA-based
system over its orthogonal multiple access (OMA) based counterpart.
More interestingly, our results show that NOMA helps in maintaining
the secrecy rate of the strong user while significantly enhancing
the secrecy performance of the weak user as compared to OMA. The correctness
of the proposed investigation is corroborated through Monte Carlo
simulation.
\end{abstract}

\section{Introduction}

\sloppy\allowdisplaybreaks

Non-orthogonal multiple access (NOMA) has gained tremendous attention
as a potential multiple access technology for the next-generation
wireless systems. It is proven to be capable of providing massive
connectivity, low latency and higher achievable rate as compared to
the traditional orthogonal multiple access (OMA) system~\cite{NOMA_book}.
In order to serve multiple users using a given resource block (time
slot, frequency band, and/or spreading code), NOMA uses power-division
multiplexing at the transmitter's side and successive interference
cancellation (SIC) at the receivers' side. Underlay spectrum sharing
is another potential technology to mitigate the problem of spectrum
scarcity, where an unlicensed/secondary network simultaneously uses
the spectrum owned by a licensed/primary network in such a manner
that the interference inflicted by the secondary network on the primary
one remains below a certain threshold~\cite{SpectrumGridlock}. The
benefits of underlay spectrum sharing NOMA systems over their OMA-based
counterparts were discussed in~\cite{CogNOMA,Kumar_SS_TCOM,Kumar_Ding_Flanagan}.

In today's privacy-concerned society, the broadcast nature of radio
waves poses a significant risk that wireless communications may be
intercepted by an eavesdropper. Therefore, physical-layer security
(PLS), which provides an additional layer of security from an information-theoretic
perspective, has gained significant attention in the past couple of
decades~\cite{PLS_Magazine}. The secrecy outage probability (SOP)
analysis of a multiple-relay-assisted two-user downlink NOMA system
was presented in~\cite{Secrecy_Ansari}, where the authors proposed
different relay selection schemes to enhance the secrecy performance
of the system. The ergodic sum secrecy rate (ESSR) analysis of a two-user
downlink and uplink cooperative NOMA system with untrusted relaying
was performed in~\cite{Lv_Secrecy}, while the authors in~\cite{Ding_Secrecy}
presented the ergodic secrecy rate (ESR) and SOP analysis of a full-duplex
relay-assisted two-user downlink NOMA system with energy harvesting.

On the other hand, some recent contributions on the secrecy analysis
of underlay spectrum sharing systems include~\cite{Secrecy_MobileComputing,Kundu_Backhaul,mMIMO_Secrecy}.
Specifically, the secrecy throughput and energy efficiency analysis
of a spectrum sharing system consisting of one primary source-destination
pair, multiple secondary source-destination pairs and an eavesdropper
was presented in~\cite{Secrecy_MobileComputing}, where primary and
secondary networks either interfere or cooperate with each other in
order to improve the secrecy performance. In~\cite{Kundu_Backhaul},
the authors derived closed-form expressions for the non-zero secrecy
rate, ESR and SOP of an outage-constrained spectrum sharing system
with transmitter selection and unreliable backhaul. The analysis of
the achievable secrecy rate for an underlay spectrum sharing multi-user
massive multiple-input multiple-output (mMIMO) system was presented
in~\cite{mMIMO_Secrecy}. However, to the best of the authors' knowledge,
the secrecy rate analysis of an underlay spectrum sharing system where
the secondary transmitter communicates with the secondary receivers
using NOMA has not yet been investigated in the literature. Therefore,
in this paper, we present the ESSR analysis of a two-user downlink
NOMA system in an underlay spectrum sharing scenario. The main contributions
in this paper are listed below:
\begin{itemize}
\item We derive an exact closed-form expression for the ESSR of a downlink
NOMA system in an underlay spectrum sharing scenario with multiple
primary receivers (PRs) considering the effect of imperfect channel
state information (CSI). We show the effect of the peak tolerable
interference power at the PRs, the maximum power budget at the secondary
transmitter (ST), channel estimation error and the number of PRs on
the ESSR of the NOMA system. 
\item We derive an asymptotic expression for the ESSR of the downlink NOMA
system for large values of peak tolerable interference at the PRs.
The asymptotic analysis confirms that the slope of the ESSR tends
to zero (w.r.t. the peak tolerable interference at the PRs), and the
ESSR becomes independent of the number of PRs as well as the quality
of the channel between the ST and PRs.
\item Using the exact closed-form expressions for the ESR, we perform a
one-dimensional numerical search to allocate power between the NOMA
users in such a manner that the ESR of the strong user in the NOMA
system and the corresponding OMA system becomes equal, and then the
remaining power is allocated to the weak NOMA user. Such a power allocation
policy ensures that the secrecy performance of the strong user in
the NOMA system is the same as that of the strong user in the corresponding
OMA system while the secrecy performance of the weak user in the NOMA
system is enhanced significantly as compared to that in the OMA system,
resulting in a performance superiority of the NOMA system over its
OMA-based counterpart in terms of ESSR.
\end{itemize}

\section{System Model}

\begin{figure}[t]
\begin{centering}
\includegraphics[width=0.98\columnwidth]{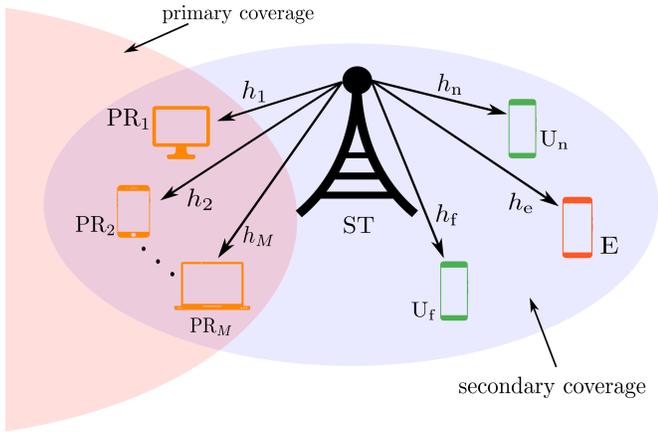}
\par\end{centering}
\caption{System model for underlay spectrum sharing NOMA with multiple primary
receivers.}

\label{fig:SysMod}
\end{figure}

Consider an underlay spectrum sharing system as shown in~Fig.~\ref{fig:SysMod},
consisting of $M$ PRs $\left(\mathrm{PR_{1},PR_{2},\ldots,PR}_{M}\right)$,
one ST, two\footnote{As suggested by the 3GPP-LTE, one resource block should be allocated
for two NOMA users. Therefore, in this paper, we consider only two
downlink NOMA users. Similar assumptions were considered in~\cite{Secrecy_Ansari,Lv_Secrecy,Ding_Secrecy}.
However, our analysis can be extended to multiple downlink NOMA users
by adopting the user pairing scheme proposed in~\cite{Ding_Secrecy}.} secondary (NOMA) receivers, denoted by $\mathrm{U_{n}}$ (near user)
and $\mathrm{U_{f}}$ (far user), and one eavesdropper $\mathrm{E}$.
It is assumed that all nodes are single-antenna devices. The distances
of $\mathrm{PR}_{m}\left(m\in\left\{ 1,2,\ldots,M\right\} \right)$,
$\mathrm{U_{n}}$, $\mathrm{U_{f}}$ and $\mathrm{E}$ from the ST
are denoted by $d_{m},d_{\mathrm{n}},d_{\mathrm{f}}$ and $d_{\mathrm{e}}$,
respectively. The channel fading coefficient for ST-$\mathrm{PR}_{m}$,
ST-$\mathrm{U_{n}}$, ST-$\mathrm{U_{f}}$ and ST-$\mathrm{E}$ links
are, respectively, given by $h_{m}\sim\mathcal{CN}\left(0,\Omega_{m}=d_{m}^{-\alpha}\right)$,
$h_{\mathrm{n}}\sim\mathcal{CN}\left(0,\Omega_{\mathrm{n}}=d_{\mathrm{n}}^{-\alpha}\right)$,
$h_{\mathrm{f}}\sim\mathcal{CN}\left(0,\Omega_{\mathrm{f}}=d_{\mathrm{f}}^{-\alpha}\right)$
and $h_{\mathrm{e}}\sim\mathcal{CN}\left(0,\Omega_{\mathrm{e}}=d_{\mathrm{e}}^{-\alpha}\right)$,
with $\alpha$ denoting the path loss exponent. The corresponding
channel gain is given by $g_{i}\triangleq\left|h_{i}\right|^{2},i\in\left\{ 1,2,\ldots,M,\mathrm{n,f,e}\right\} $.
We assume that the ST has imperfect instantaneous CSI of the ST-$\mathrm{PR}_{m}$,
ST-$\mathrm{U_{n}}$, ST-$\mathrm{U_{f}}$ and ST-$\mathrm{E}$ links.
Let the estimate of $h_{i}$ be denoted by $\tilde{h}_{i}$, and $h_{i}=\tilde{h}_{i}+\epsilon_{i}$,
where $\epsilon_{i}\sim\mathcal{CN}\left(0,\sigma_{\epsilon}^{2}\right)$
is the channel estimation error. Therefore, $\tilde{h}_{i}\sim\mathcal{CN}\left(0,\tilde{\Omega}_{i}\right)$,
where $\tilde{\Omega}_{i}=\Omega_{i}-\sigma_{\epsilon}^{2}$ and the
corresponding channel gain is denoted by $\tilde{g}_{i}$. The peak
(instantaneous) interference that the PRs can tolerate from the secondary
network is denoted by $I_{\mathrm{p}}$, and the maximum transmit
power at the ST is denoted by $P_{\max}$. Therefore, the instantaneous
transmit power of the ST is given by $P=\min\left\{ P_{\max},\tfrac{I_{\mathrm{p}}}{\tilde{g}_{1}},\tfrac{I_{\mathrm{p}}}{\tilde{g}_{2}},\ldots,\tfrac{I_{\mathrm{p}}}{\tilde{g}_{M}}\right\} =\min\left\{ P_{\max},\tfrac{I_{\mathrm{p}}}{\tilde{g}_{\mathrm{p}}}\right\} $,
where $\tilde{g}_{\mathrm{p}}\triangleq\max_{m}\tilde{g}_{m}$. We
denote the probability density function (PDF), cumulative distribution
function (CDF) and complementary CDF (CCDF) of a random variable $\mathcal{X}$
by $f_{\mathcal{X}}\left(\cdot\right)$, $F_{\mathcal{X}}\left(\cdot\right)$
and $\mathcal{F_{X}}\left(\cdot\right)$, respectively, and we define
$\mathscr{F}_{\mathcal{X}}\left(c_{1},c_{2}\right)\triangleq\int_{c_{1}}^{c_{2}}f_{\mathcal{X}}\left(t\right)\mathrm{d}t$.

\section{Analysis of the ESSR}

Based on the estimated CSI at the ST, the users $\mathrm{U_{n}}$
and $\mathrm{U_{f}}$ are further categorized as $\mathrm{U_{s}}$
(strong user) and $\mathrm{U_{w}}$ (weak user), where $\mathrm{s}\triangleq\mathrm{argmax}_{j\in\left\{ \mathrm{n,f}\right\} }\ \tilde{g}_{j}$
and $\mathrm{w}\triangleq\mathrm{argmin}_{j\in\left\{ \mathrm{n,f}\right\} }\ \tilde{g}_{j}$.
The signal received at $\mathrm{U}_{k},k\in\left\{ \mathrm{s},\mathrm{w}\right\} $
is given by 
\[
y_{k}=\left(\tilde{h}_{k}+\epsilon_{k}\right)\left(\sqrt{a_{\mathrm{s}}P}x_{\mathrm{s}}+\sqrt{a_{\mathrm{w}}P}x_{\mathrm{w}}\right)+n_{k},
\]
where $x_{k}$ is the unit-energy information-bearing complex constellation
symbol and $n_{k}\sim\mathcal{CN}\left(0,1\right)$ is the additive
white Gaussian noise (AWGN) at node $\mathrm{U}_{k}$. Following the
NOMA principle, it is assumed that $a_{\mathrm{s}}<a_{\mathrm{w}}$
and $a_{\mathrm{s}}+a_{\mathrm{w}}\mathrm{=1}$. The weak user decodes
$x_{\mathrm{w}}$ by treating the interference due to $x_{\mathrm{s}}$
as additional noise, whereas the strong user first decodes $x_{\mathrm{w}}$
by treating the interference due to $x_{\mathrm{s}}$ as additional
noise and then applies successive interference cancellation (SIC)
to decode $x_{\mathrm{s}}$. Therefore, the signal-to-interference-plus-noise
ratio (SINR) at $\mathrm{U_{s}}$ and $\mathrm{U_{w}}$ are, respectively,
given by 
\[
\gamma_{\mathrm{s}}=\dfrac{a_{\mathrm{s}}\tilde{g}_{\mathrm{s}}P}{1+\sigma_{\epsilon}^{2}P},\qquad\gamma_{\mathrm{w}}=\dfrac{a_{\mathrm{w}}\tilde{g}_{\mathrm{w}}P}{1+a_{\mathrm{s}}\tilde{g}_{\mathrm{w}}P+\sigma_{\epsilon}^{2}P}.
\]
Following~\cite[Ch. 15]{Secrecy_Book}, we consider the case that
$\mathrm{E}$ follows the same decoding order as that of the legitimate
users. Therefore, the SINR at $\mathrm{E}$ for decoding $x_{\mathrm{w}}$
and $x_{\mathrm{s}}$ are respectively given by
\[
\gamma_{\mathrm{e,w}}=\dfrac{a_{\mathrm{w}}\tilde{g}_{\mathrm{e}}P}{1+a_{\mathrm{s}}\tilde{g}_{\mathrm{e}}P+\sigma_{\epsilon}^{2}P},\qquad\gamma_{\mathrm{e,s}}=\dfrac{a_{\mathrm{s}}\tilde{g}_{\mathrm{e}}P}{1+\sigma_{\epsilon}^{2}P}.
\]
Let $\hat{g}_{\mathrm{s}}\triangleq\sigma_{\epsilon}^{2}+a_{\mathrm{s}}\tilde{g}_{\mathrm{s}}$,
$\hat{g}_{\mathrm{w}}\triangleq\sigma_{\epsilon}^{2}+a_{\mathrm{s}}\tilde{g}_{\mathrm{w}}$,
$\check{g}_{\mathrm{w}}\triangleq\sigma_{\epsilon}^{2}+\tilde{g}_{\mathrm{w}}$
and $\hat{g}_{\mathrm{e}}\triangleq\sigma_{\epsilon}^{2}+a_{\mathrm{s}}\tilde{g}_{\mathrm{e}}$.
Then using the standard statistical procedure of transformation of
random variables, it follows that
\[
f_{\hat{g}_{\mathrm{s}}}\!\left(x\right)\!=\!\!\!\sum_{\ell\in\left\{ 1,2,3\right\} }\!\!\!\!\!\!\hat{A}_{\ell}\exp\!\left(\!\dfrac{-x}{a_{\mathrm{s}}\Xi_{\ell}}\!\right),\mathcal{F}_{\hat{g}_{\mathrm{s}}}\!\left(x\right)\!=\!\!\!\sum_{\ell\in\left\{ 1,2,3\right\} }\!\!\!\!\!\!\hat{\mathcal{A}}_{\ell}\exp\!\left(\!\dfrac{-x}{a_{\mathrm{s}}\Xi_{\ell}}\!\right),
\]

\[
f_{\hat{g}_{\mathrm{w}}}\left(x\right)=-\hat{A}_{2}\exp\left(\dfrac{-x}{a_{\mathrm{s}}\Xi_{2}}\right),\mathcal{F}_{\hat{g}_{\mathrm{w}}}\left(x\right)=-\hat{\mathcal{A}}_{2}\exp\left(\dfrac{-x}{a_{\mathrm{s}}\Xi_{2}}\right),
\]
\[
f_{\check{g}_{\mathrm{w}}}\left(x\right)=\check{A}_{2}\exp\left(\dfrac{-x}{\Xi_{2}}\right),\mathcal{F}_{\check{g}_{\mathrm{w}}}\left(x\right)=\check{\mathcal{A}}_{2}\exp\left(\dfrac{-x}{\Xi_{2}}\right),
\]

\[
f_{\hat{g}_{\mathrm{e}}}\!\left(x\right)\!=\!\hat{C}_{\mathrm{e}}\exp\left(\!\dfrac{-x}{a_{\mathrm{s}}\tilde{\Omega}_{\mathrm{e}}}\!\right),\mathscr{F}_{\hat{g}_{\mathrm{e}}}\!\left(\sigma_{\epsilon}^{2},x\right)\!=1-\hat{\mathscr{C}}_{\mathrm{e}}\exp\left(\!\dfrac{-x}{a_{\mathrm{s}}\tilde{\Omega}_{\mathrm{e}}}\!\right),
\]
\[
\mathscr{F}_{\check{g}_{\mathrm{e}}}\left(\sigma_{\epsilon}^{2},x\right)=1-\check{\mathscr{C}}_{\mathrm{e}}\exp\left(\dfrac{-x}{\tilde{\Omega}_{\mathrm{e}}}\right),
\]
where $\Xi_{1}=\tilde{\Omega}_{\mathrm{n}}$, $\Xi_{2}=\tilde{\Omega}_{\mathrm{n,f}}$,
$\Xi_{3}=\tilde{\Omega}_{\mathrm{f}}$, $\tilde{\Omega}_{\mathrm{n,f}}=\left(\tilde{\Omega}_{\mathrm{n}}^{-1}+\tilde{\Omega}_{\mathrm{f}}^{-1}\right)^{-1}$,
$\hat{A}_{\ell}=\tfrac{(-1)^{\ell+1}}{a_{\mathrm{s}}\Xi_{\ell}}\exp\left(\tfrac{\sigma_{\epsilon}^{2}}{a_{\mathrm{s}}\Xi_{\ell}}\right)$,
$\hat{\mathcal{A}}_{\ell}=a_{\mathrm{s}}\Xi_{\ell}\hat{A}_{\ell}$,
$\check{A}_{2}=\tfrac{1}{\Xi_{2}}\exp\left(\tfrac{\sigma_{\epsilon}^{2}}{\Xi_{2}}\right)$,
$\check{\mathcal{A}}_{2}=\Xi_{2}\check{A}_{2}$, $\hat{C}_{\mathrm{e}}=\tfrac{1}{a_{\mathrm{s}}\tilde{\Omega}_{\mathrm{e}}}\exp\left(\tfrac{\sigma_{\epsilon}^{2}}{a_{\mathrm{s}}\tilde{\Omega}_{\mathrm{e}}}\right)$
and $\hat{\mathscr{C}}_{\mathrm{e}}=a_{\mathrm{s}}\tilde{\Omega}_{\mathrm{e}}\hat{C}_{\mathrm{e}}$.
From~\cite{CostaPDF}, the PDF of $\tilde{g}_{\mathrm{p}}$ is given
by 
\[
f_{\tilde{g}_{\mathrm{p}}}\left(x\right)=-\sum_{\boldsymbol{\eta}\in\Phi}\kappa_{\boldsymbol{\eta}}B_{\boldsymbol{\eta}}\exp\left(-B_{\boldsymbol{\eta}}x\right),
\]
where $\Phi=\left\{ \boldsymbol{\eta}=[\eta_{1}\eta_{2}\cdots\eta_{M}]\in\mathbb{Z}_{2}^{M}\left|\sum\nolimits _{m=1}^{M}\eta_{m}>0\right.\right\} $
represents the set of all nonzero binary vectors of length $M$, $\mathbb{Z}_{2}=\left\{ 0,1\right\} $,
$\kappa_{\boldsymbol{\eta}}=\prod_{m=1}^{M}(-1)^{\eta_{m}}$ and $B_{\boldsymbol{\eta}}=\sum_{m=1}^{M}\left(\eta_{m}/\tilde{\Omega}_{m}\right)$. 

Finally, the ESR of the strong user can be written as 
\begin{align}
\bar{R}_{\mathrm{s}} & =\mathbb{E}\left\{ \max\left\{ \log_{2}\left(1+\gamma_{\mathrm{s}}\right)-\log_{2}\left(1+\gamma_{\mathrm{e,s}}\right),0\right\} \right\} \nonumber \\
 & =\mathbb{E}\left\{ \max\left\{ \log_{2}\left(1+\hat{g}_{\mathrm{s}}P\right)-\log_{2}\left(1+\hat{g}_{\mathrm{e}}P\right),0\right\} \right\} .\label{eq:RsAvgDef}
\end{align}

\begin{thm}
\label{thm:RsAvg}A closed-form expression for the ESR of $\mathrm{U_{s}}$
can be derived as
\begin{equation}
\bar{R}_{\mathrm{s}}=\mathscr{I}_{\mathrm{s,1}}+\mathscr{I}_{\mathrm{s,2}}-\mathscr{I}_{\mathrm{s,3}}-\mathscr{I}_{\mathrm{s,4}},\label{eq:RsClosed}
\end{equation}
where the expressions for $\mathscr{I}_{\mathrm{s,1}},\mathscr{I}_{\mathrm{s,2}},\mathscr{I}_{\mathrm{s,3}}$
and $\mathscr{I}_{\mathrm{s,4}}$ are given by~\eqref{eq:Is1Closed}
\textendash{} \eqref{eq:Is4Closed}, shown at the top of the next
page, $\Theta\triangleq I_{\mathrm{p}}/P_{\max}$, $\Xi_{\ell,e}\triangleq\left(\Xi_{\ell}^{-1}+\tilde{\Omega}^{-1}\right)^{-1}$,
and $\mathrm{Ei}\left(\cdot\right)$ and $\mathrm{E_{1}}\left(\cdot\right)$
are the exponential integrals.

\begin{figure*}[t]
\begin{equation}
\mathscr{I}_{\mathrm{s,1}}=-\log_{2}\left(e\right)a_{\mathrm{s}}\text{\ensuremath{\left[\sum_{\ell\in\left\{  1,2,3\right\}  }\hat{A}_{\ell}\left\{  \Xi_{\ell}\mathbb{J}_{1}\left(a_{\mathrm{s}}\Xi_{\ell}\right)-\hat{\mathscr{C}}_{\mathrm{e}}\Xi_{\ell,\mathrm{e}}\mathbb{J}_{1}\left(a_{\mathrm{s}}\Xi_{\ell,\mathrm{e}}\right)\right\}  \right]}}\left[\sum_{\boldsymbol{\eta}\in\Phi}\kappa_{\boldsymbol{\eta}}\left\{ 1-\exp\left(-B_{\boldsymbol{\eta}}\Theta\right)\right\} \right],\label{eq:Is1Closed}
\end{equation}
\begin{equation}
\mathscr{I}_{\mathrm{s,2}}=-\log_{2}\left(e\right)a_{\mathrm{s}}\sum_{\ell\in\left\{ 1,2,3\right\} }\sum_{\boldsymbol{\eta}\in\Phi}\hat{A}_{\ell}\kappa_{\eta}\left[\Xi_{\ell}\mathbb{J}_{2}\left(a_{\mathrm{s}}\Xi_{\ell}\right)-\hat{\mathscr{C}}_{\mathrm{e}}\Xi_{\ell,\mathrm{e}}\mathbb{J}_{2}\left(a_{\mathrm{s}}\Xi_{\ell,\mathrm{e}}\right)\right],\label{eq:Is2Closed}
\end{equation}
\begin{equation}
\mathscr{I}_{\mathrm{s,3}}=-\log_{2}\left(e\right)a_{\mathrm{s}}\hat{C}_{\mathrm{e}}\left[\sum_{\ell\in\left\{ 1,2,3\right\} }\hat{\mathcal{A}}_{\ell}\Xi_{\ell,\mathrm{e}}\mathbb{J}_{1}\left(a_{\mathrm{s}}\Xi_{\ell,\mathrm{e}}\right)\right]\left[\sum_{\boldsymbol{\eta}\in\Phi}\kappa_{\boldsymbol{\eta}}\left\{ 1-\exp\left(-B_{\boldsymbol{\eta}}\Theta\right)\right\} \right],\label{eq:Is3Closed}
\end{equation}
\begin{equation}
\mathscr{I}_{\mathrm{s,4}}=-\log_{2}\left(e\right)a_{\mathrm{s}}\hat{C}_{\mathrm{e}}\sum_{\ell\in\left\{ 1,2,3\right\} }\sum_{\boldsymbol{\eta}\in\Phi}\hat{\mathcal{A}}_{\ell}\kappa_{\boldsymbol{\eta}}\Xi_{\ell,2}\mathbb{J}_{2}\left(a_{\mathrm{s}}\Xi_{\ell,\mathrm{e}}\right),\label{eq:Is4Closed}
\end{equation}
\begin{equation}
\mathbb{J}_{1}\left(\Omega\right)=\exp\left(\dfrac{-\sigma_{\epsilon}^{2}}{\Omega}\right)\ln\left(1+\sigma_{\epsilon}^{2}P_{\max}\right)+\exp\left(\dfrac{1}{\Omega P_{\max}}\right)\mathrm{E}_{1}\left[\dfrac{1}{\Omega}\left(\dfrac{1}{P_{\max}}+\sigma_{\epsilon}^{2}\right)\right],\label{eq:J1Closed}
\end{equation}
\begin{align}
\mathbb{J}_{2}\left(\Omega\right)\triangleq & \exp\left(\dfrac{-\sigma_{\epsilon}^{2}}{\Omega}\right)\mathrm{Ei}\left(-B_{\boldsymbol{\eta}}\Theta\right)-\dfrac{B_{\boldsymbol{\eta}}^{-1}}{B_{\boldsymbol{\eta}}^{-1}-I_{\mathrm{p}}\Omega}\left\{ \exp\left(I_{\mathrm{p}}\sigma_{\epsilon}^{2}B_{\boldsymbol{\eta}}-\dfrac{\sigma_{\epsilon}^{2}}{\Omega}\right)\mathrm{Ei}\left(-B_{\boldsymbol{\eta}}\left(\Theta+I_{\mathrm{p}}\sigma_{\epsilon}^{2}\right)\right)-\exp\left(\dfrac{\Theta}{I_{\mathrm{p}}\Omega}-B_{\boldsymbol{\eta}}\Theta\right)\right.\nonumber \\
 & \left.\times\mathrm{Ei}\left(\dfrac{-\Theta-I_{\mathrm{p}}\sigma_{\epsilon}^{2}}{I_{\mathrm{p}}\Omega}\right)\right\} +\exp\left(-B_{\boldsymbol{\eta}}\Theta-\dfrac{\sigma_{\epsilon}^{2}}{\Omega}\right)\left\{ \ln\left(1+\dfrac{I_{\mathrm{p}}\sigma_{\epsilon}^{2}}{\Theta}\right)+\exp\left(\dfrac{\Theta+I_{\mathrm{p}}\sigma_{\epsilon}^{2}}{I_{\mathrm{p}}\Omega}\right)\mathrm{E}_{1}\left(\dfrac{\Theta+I_{\mathrm{p}}\sigma_{\epsilon}^{2}}{I_{\mathrm{p}}\Omega}\right)\right\} .\label{eq:J2Closed}
\end{align}

\rule{2.05\columnwidth}{0.5pt}
\end{figure*}
 
\end{thm}
\begin{IEEEproof}
See Appendix~\ref{sec:Proof-of-Theorem-RsAvg}.
\end{IEEEproof}
On the other hand, the ESR for the weak user $\mathrm{U_{w}}$ can
be written as
\begin{align}
 & \bar{R}_{\mathrm{w}}=\mathbb{E}\left\{ \max\left\{ \log_{2}\left(1+\gamma_{\mathrm{w}}\right)-\log_{2}\left(1+\gamma_{\mathrm{e,w}}\right),0\right\} \right\} \nonumber \\
 & =\mathbb{E}\left\{ \max\left\{ \log_{2}\left(\dfrac{1+\left(\tilde{g}_{_{\mathrm{w}}}+\sigma_{\epsilon}^{2}\right)P}{1+\left(a_{\mathrm{s}}\tilde{g}_{\mathrm{_{\mathrm{w}}}}+\sigma_{\epsilon}^{2}\right)P}\right)\right.\right.\nonumber \\
 & \qquad\qquad\qquad-\left.\left.\log_{2}\left(\dfrac{1+\left(\tilde{g}_{_{\mathrm{e}}}+\sigma_{\epsilon}^{2}\right)P}{1+\left(a_{\mathrm{s}}\tilde{g}_{_{\mathrm{e}}}+\sigma_{\epsilon}^{2}\right)P}\right),0\right\} \right\} .\label{eq:RwAvgDef}
\end{align}

\begin{thm}
\label{thm:RwAvg} A closed-form expression for the ESR of $\mathrm{U_{w}}$
can be derived as 
\begin{equation}
\bar{R}_{\mathrm{w}}=\mathscr{I}_{\mathrm{w,1}}+\mathscr{I}_{\mathrm{w,2}}-\mathscr{I}_{\mathrm{w,3}}-\mathscr{I}_{\mathrm{w,4}},\label{eq:RwClosed}
\end{equation}
where the expressions for $\mathscr{I}_{\mathrm{w,1}},\mathscr{I}_{\mathrm{w,2}},\mathscr{I}_{\mathrm{w,3}}$
and $\mathscr{I}_{\mathrm{w,4}}$ are given by~\eqref{eq:Iw1Closed}
\textendash{} \eqref{eq:Iw4Closed}, shown on the next page. 

\begin{figure*}[t]
\begin{equation}
\mathscr{I}_{\mathrm{w,1}}\!=\!-\log_{2}\left(e\right)\!\sum_{\boldsymbol{\eta}\in\Phi}\!\kappa_{\boldsymbol{\eta}}\left\{ 1\!-\!\exp\left(\!-B_{\boldsymbol{\eta}}\Theta\right)\right\} \!\!\left[\!\check{A}_{2}\Xi_{2}\mathbb{J}_{1}\!\left(\Xi_{2}\right)\!-\!\check{A}_{2}\check{\mathscr{C}}_{\mathrm{e}}\Xi_{2,\mathrm{e}}\mathbb{J}_{1}\!\left(\Xi_{2,\mathrm{e}}\right)\!+\!\hat{A}_{2}a_{\mathrm{s}}\Xi_{2}\mathbb{J}_{1}\!\left(a_{\mathrm{s}}\Xi_{2}\right)\!-\!\hat{A}_{2}\hat{\mathscr{C}}_{\mathrm{e}}a_{\mathrm{s}}\Xi_{2,\mathrm{e}}\mathbb{J}_{1}\!\left(a_{\mathrm{s}}\Xi_{2,\mathrm{e}}\right)\right],\label{eq:Iw1Closed}
\end{equation}
\begin{equation}
\mathscr{I}_{\mathrm{w,2}}=\!-\log_{2}\left(e\right)\sum_{\boldsymbol{\eta}\in\Phi}\kappa_{\boldsymbol{\eta}}\left[\check{A}_{2}\Xi_{2}\mathbb{J}_{2}\left(\Xi_{2}\right)\!-\!\check{A}_{2}\check{\mathscr{C}}_{\mathrm{e}}\Xi_{2,\mathrm{e}}\mathbb{J}_{2}\left(\Xi_{2,\mathrm{e}}\right)+\hat{A}_{2}a_{\mathrm{s}}\Xi_{2}\mathbb{J}_{2}\left(a_{\mathrm{s}}\Xi_{2}\right)-\hat{A}_{2}\hat{\mathscr{C}}_{\mathrm{e}}a_{\mathrm{s}}\Xi_{2,\mathrm{e}}\mathbb{J}_{2}\left(a_{\mathrm{s}}\Xi_{2,\mathrm{e}}\right)\right],\label{eq:Iw2Closed}
\end{equation}
\begin{equation}
\mathscr{I}_{\mathrm{w,3}}=-\log_{2}\left(e\right)\sum_{\boldsymbol{\eta}\in\Phi}\kappa_{\boldsymbol{\eta}}\left\{ 1-\exp\left(-B_{\boldsymbol{\eta}}\Theta\right)\right\} \left[\check{\mathcal{A}}_{2}\check{C}_{\mathrm{e}}\Xi_{2,\mathrm{e}}\mathbb{J}_{1}\left(\Xi_{2,\mathrm{e}}\right)+\hat{\mathcal{A}}_{2}\hat{C}_{\mathrm{e}}a_{\mathrm{s}}\Xi_{2,\mathrm{e}}\mathbb{J}_{1}\left(a_{\mathrm{s}}\Xi_{2,\mathrm{e}}\right)\right],\label{eq:Iw3Closed}
\end{equation}
\begin{equation}
\mathscr{I}_{\mathrm{w,4}}=-\log_{2}\left(e\right)\sum_{\boldsymbol{\eta}\in\Phi}\kappa_{\boldsymbol{\eta}}\left[\check{\mathcal{A}}_{2}\check{C}_{\mathrm{e}}\Xi_{2,\mathrm{e}}\mathbb{J}_{2}\left(\Xi_{2,\mathrm{e}}\right)+a_{\mathrm{s}}\hat{\mathcal{A}}_{2}\hat{C}_{\mathrm{e}}\Xi_{2,\mathrm{e}}\mathbb{J}_{2}\left(a_{\mathrm{s}}\Xi_{2,\mathrm{e}}\right)\right].\label{eq:Iw4Closed}
\end{equation}
\rule{2.05\columnwidth}{0.5pt}
\end{figure*}
\end{thm}
\begin{IEEEproof}
See Appendix~\ref{sec:Proof-of-Theorem-RwAvg}.
\end{IEEEproof}
Using~\eqref{eq:RsClosed} and~\eqref{eq:RwClosed}, a closed-form
expression for the ESSR of the NOMA system is given by $\bar{R}_{\mathrm{sum}}=\bar{R}_{\mathrm{s}}+\bar{R}_{\mathrm{w}}$. 

\paragraph*{ESSR for OMA}

For the case of an underlay spectrum sharing OMA system, the ST transmits
$\sqrt{P}x_{\mathrm{s}}$ to $\mathrm{U_{s}}$ in the first time slot,
and $\sqrt{P}x_{\mathrm{w}}$ to $\mathrm{U_{w}}$ in the second time
slot. For a fair comparison between the NOMA and OMA systems, we consider
the same value of $I_{\mathrm{p}}$ for both systems. Therefore, the
ESSR for the OMA system is given by
\begin{align}
 & \bar{R}_{\mathrm{sum,OMA}}=\bar{R}_{\mathrm{s,OMA}}+\bar{R}_{\mathrm{w,OMA}}\nonumber \\
 & =\sum_{k\in\left\{ \mathrm{s,w}\right\} }\mathbb{E}\left\{ 0.5\max\left\{ \log_{2}\left(\dfrac{1+\tilde{g}_{k}P}{1+\tilde{g}_{\mathrm{e}}P}\right),0\right\} \right\} .\label{eq:RsumOMADef}
\end{align}
As the focus of this paper is on the NOMA-based system, we do not
provide the closed-form expression for $\bar{R}_{\mathrm{sum,OMA}}$.
Next, we carry out the asymptotic analysis $\left(I_{\mathrm{p}}\to\infty\right)$
of the ESSR for the NOMA system.

\section{Asymptotic ESSR for the NOMA System}
\begin{thm}
An analytical expression for the asymptotic $\left(I_{\mathrm{p}}\to\infty\right)$
ESSR for the NOMA system can be derived as 
\begin{equation}
\bar{R}_{\mathrm{sum,asymp}}=\mathcal{I}_{\mathrm{s}}+\mathcal{I}_{\mathrm{w}},\label{eq:RsumAsympClosed}
\end{equation}
where the closed-form expressions for $\mathcal{I}_{\mathrm{s}}$
and $\mathcal{I}_{\mathrm{w}}$ are given by~\eqref{eq:mathcal-Is-Closed}
and~\eqref{eq:mathcal-Iw-Closed}, respectively, shown on the next
page.

\begin{figure*}[t]
\begin{align}
 & \mathcal{I}_{\mathrm{s}}=\log_{2}\left(e\right)a_{\mathrm{s}}\sum_{\ell\in\left\{ 1,2,3\right\} }\left[\hat{A}_{\ell}\Xi_{\ell}\mathbb{J}_{1}\left(a_{\mathrm{s}}\Xi_{\ell}\right)-\hat{A}_{\ell}\hat{\mathscr{C}}_{\mathrm{e}}\Xi_{\ell,\mathrm{e}}\mathbb{J}_{1}\left(a_{\mathrm{s}}\Xi_{\ell,\mathrm{e}}\right)-\hat{\mathcal{A}}_{\ell}\hat{C}_{\mathrm{e}}\Xi_{\ell,\mathrm{e}}\mathbb{J}_{1}\left(a_{\mathrm{s}}\Xi_{\ell,\mathrm{e}}\right)\right],\label{eq:mathcal-Is-Closed}
\end{align}
\begin{align}
 & \mathcal{I}_{\mathrm{w}}=\log_{2}\left(e\right)\left[\check{A}_{2}\Xi_{2}\mathbb{J}_{1}\!\left(\Xi_{2}\right)\!-\!\check{A}_{2}\check{\mathscr{C}}_{\mathrm{e}}\Xi_{2,\mathrm{e}}\mathbb{J}_{1}\!\left(\Xi_{2,\mathrm{e}}\right)\!+\!\hat{A}_{2}a_{\mathrm{s}}\Xi_{2}\mathbb{J}_{1}\!\left(a_{\mathrm{s}}\Xi_{2}\right)\!-\!\hat{A}_{2}\hat{\mathscr{C}}_{\mathrm{e}}a_{\mathrm{s}}\Xi_{2,\mathrm{e}}\mathbb{J}_{1}\!\left(a_{\mathrm{s}}\Xi_{2,\mathrm{e}}\right)\right.\nonumber \\
 & \qquad\qquad\qquad\qquad\qquad\qquad\qquad\qquad\qquad\qquad\qquad\qquad-\left.\check{\mathcal{A}}_{2}\check{C}_{\mathrm{e}}\Xi_{2,\mathrm{e}}\mathbb{J}_{1}\left(\Xi_{2,\mathrm{e}}\right)-\hat{\mathcal{A}}_{2}\hat{C}_{\mathrm{e}}a_{\mathrm{s}}\Xi_{2,\mathrm{e}}\mathbb{J}_{1}\left(a_{\mathrm{s}}\Xi_{2,\mathrm{e}}\right)\right].\label{eq:mathcal-Iw-Closed}
\end{align}

\rule{2.05\columnwidth}{0.5pt}
\end{figure*}
\end{thm}
\begin{IEEEproof}
Note that for $I_{\mathrm{p}}\to\infty$ and finite $P_{\max}$, we
have $\Theta=\tfrac{I_{\mathrm{P}}}{P_{\max}}\to\infty$. Letting
$\Theta\to\infty$ in~\eqref{eq:RsAvgDef-New} and~\eqref{eq:RwAvgDef-New},
and then solving the integrals following a similar line of argument
as given in~Appendix~\ref{sec:Proof-of-Theorem-RsAvg}, a closed-form
expression for $\bar{R}_{\mathrm{sum,asymp}}$ is given by~\eqref{eq:RsumAsympClosed}.
This concludes the proof. 
\end{IEEEproof}
It can be noted from~\eqref{eq:RsumAsympClosed} \textendash{} \eqref{eq:mathcal-Iw-Closed}
that for large values of $I_{\mathrm{p}}$, the ESSR of the NOMA is
independent of the number of PRs as well as the quality of the link
between the ST and PRs. Also, the slope of the ESSR w.r.t $I_{\mathrm{p}}$
becomes equal to zero for $I_{\mathrm{p}}\to\infty$. 
\begin{figure}[t]
\begin{centering}
\includegraphics[width=0.85\columnwidth,height=5.6cm]{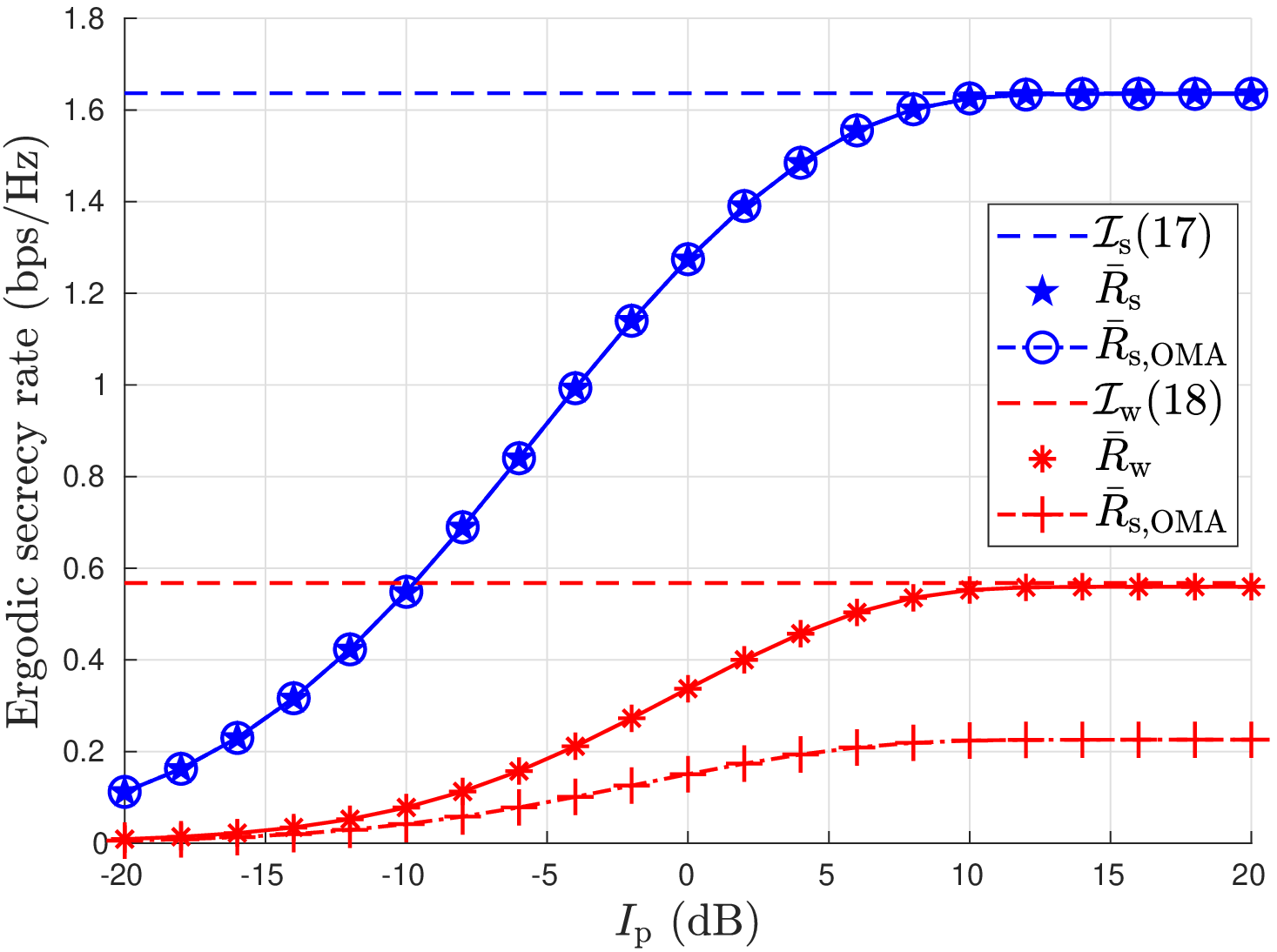}
\par\end{centering}
\caption{Comparison of ESR for NOMA and OMA systems for $P_{\max}=50$~dB.}
\label{fig:RateComponents}
\end{figure}
\begin{figure}[t]
\begin{centering}
\includegraphics[width=0.85\columnwidth,height=5.6cm]{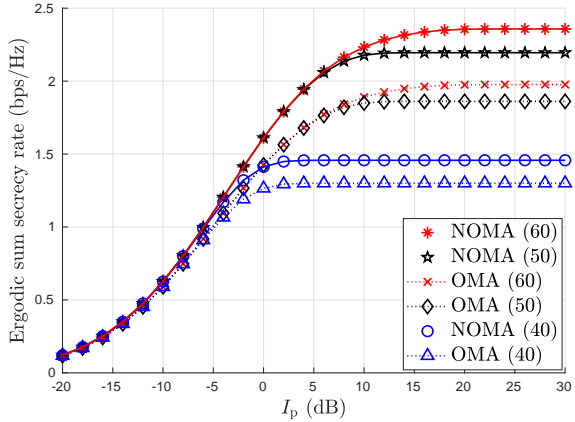}
\par\end{centering}
\caption{Comparison of ESSR for NOMA and OMA systems. The number in the parentheses
denotes $P_{\max}$ in dB.}
\label{fig:SumRate}
\end{figure}
\begin{figure}[t]
\begin{centering}
\includegraphics[width=0.85\columnwidth,height=5.6cm]{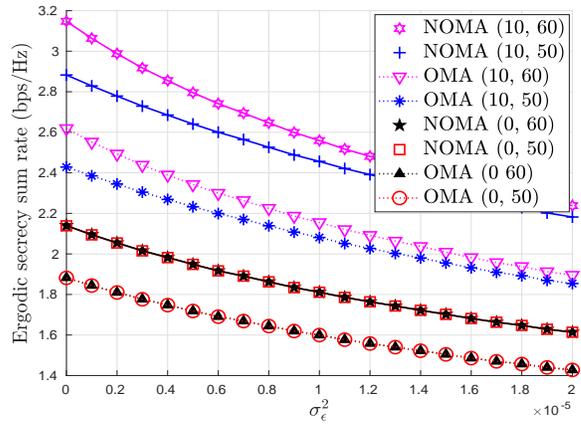}
\par\end{centering}
\caption{Effect of channel estimation error on the ESSR. The numbers in the
parentheses denote $\left(I_{\mathrm{p}},P_{\max}\right)$ in dB.}
\label{fig:SumRate_sseVar}
\end{figure}
\begin{figure}[t]
\begin{centering}
\includegraphics[width=0.85\columnwidth,height=5.6cm]{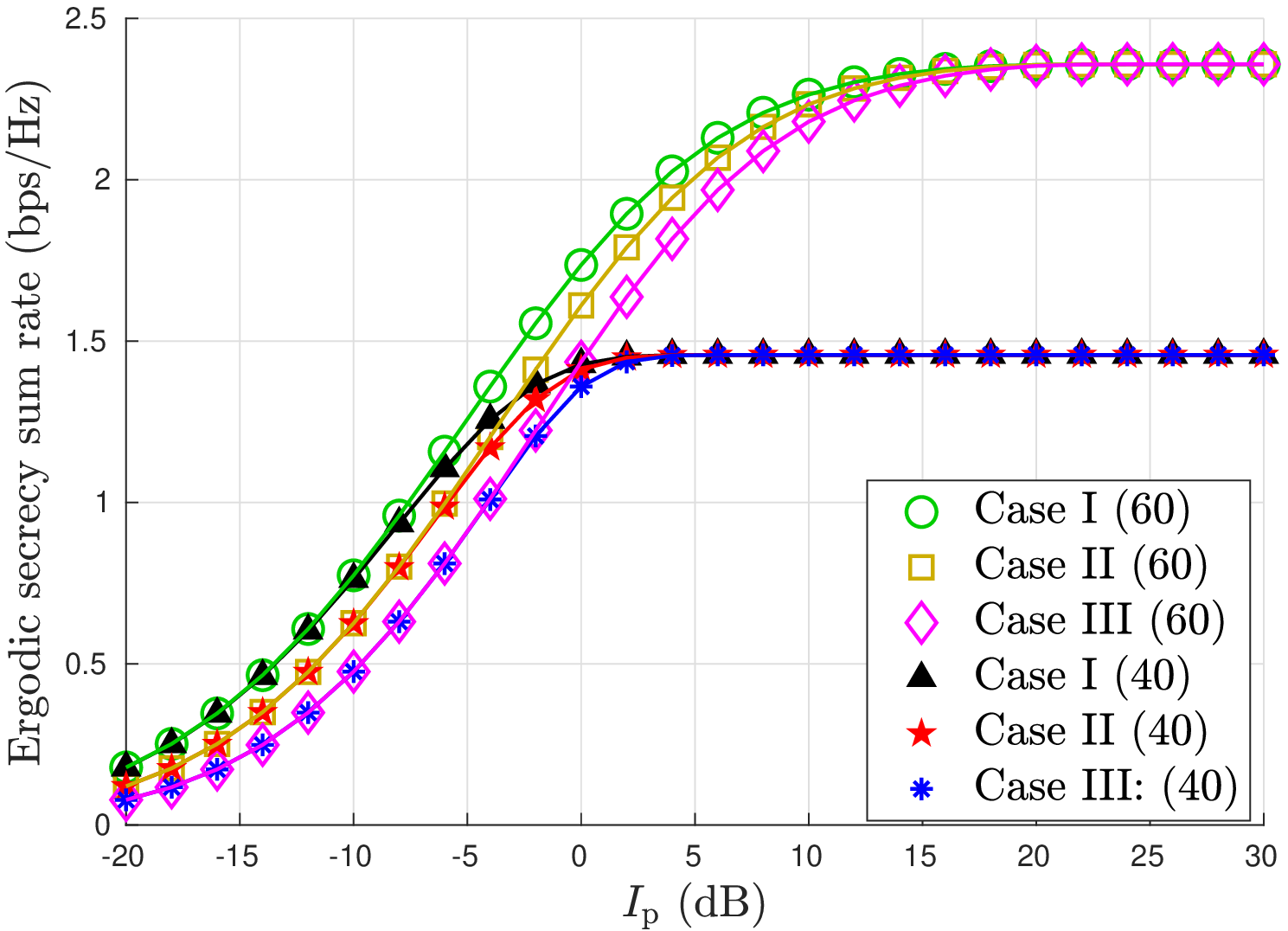}
\par\end{centering}
\caption{Effect of the number of PRs on the ESSR of the NOMA system. The number
in the parentheses denotes $P_{\max}$ in dB.}
\label{fig:SumRate_PRVar}
\end{figure}

\section{Results and Discussions}

In this section, we present the numerical and analytical results for
the ESSR of NOMA and OMA systems. We consider a system with $d_{\mathrm{n}}=30$~m,
$d_{\mathrm{f}}=100$~m, $d_{\mathrm{e}}=150$~m, $\left\{ d_{1},d_{2},d_{3},d_{4}\right\} =\left\{ 200,205,210,215\right\} $~m,
$\sigma_{\epsilon}^{2}=2\times10^{-5}$, and $\alpha=2$, unless stated
otherwise. In the case of NOMA, we use the bisection method to find
the value of $a_{\mathrm{s}}$ such that $\bar{R}_{\mathrm{s}}=\bar{R}_{\mathrm{s,OMA}}$,
and then the remaining power fraction $\left(a_{\mathrm{w}}\right)$
is allocated to the weak user. In each figure, the legend indicates
the numerically obtained results, while the solid curves indicate
the analytical results.

Fig.~\ref{fig:RateComponents} shows a performance comparison of
the ESR of the NOMA and OMA systems. It can be observed from the figure
that the ESR of $\mathrm{U_{s}}$ in the NOMA system is equal to that
in the case of the OMA system (by virtue of the power allocation scheme),
whereas the ESR of $\mathrm{U_{w}}$ in the NOMA system is higher
that that of the OMA system. This result confirms that NOMA helps
in maintaining the secrecy rate of the strong user while significantly
enhancing the secrecy performance of the weak user as compared to
OMA. We also show the asymptotic $\text{\ensuremath{\left(I_{\mathrm{p}}\to\infty\right)}}$
ESR for the NOMA system, which matches well with the exact ESR for
large $I_{\mathrm{p}}$. Moreover, a close agreement between the numerical
and analytical results (for the case of NOMA) verifies the correctness
of the derived closed-form expressions.

Fig.~\ref{fig:SumRate} shows a comparison of the ESSR for the NOMA
and OMA systems. Note that the NOMA system outperforms its OMA-based
counterpart by achieving a significantly higher ESSR. The ESSR for
the NOMA/OMA system first increases with an increase of the value
of $I_{\mathrm{p}}$ (which we refer to the \textit{interference-constrained
regime}) and then saturates for larger value of $I_{\mathrm{p}}$
(which we refer to as the \textit{power-constrained regime}). Interestingly,
in the interference-constrained regime, the ESSR of the NOMA system
remains the same irrespective of the value of $P_{\max}$, whereas,
in the power-constrained regime, the ESSR is independent of the value
of $I_{\mathrm{p}}$. A similar behavior is evident for the OMA system.
It is noteworthy that for the system parameters considered in Fig.~\ref{fig:SumRate},
$\min\left(P_{\max},\tfrac{I_{\mathrm{p}}}{\mathbb{E}\left\{ \tilde{g}_{\mathrm{p}}\right\} }\right)=\min\left(P_{\max},\tfrac{I_{\mathrm{p}}}{\tilde{\Omega}_{\mathrm{p}}}\right)$
becomes constant $\left(=P_{\max}\right)$ w.r.t $I_{p}$ at $I_{\mathrm{p}}\gtrsim0$
dB for $P_{\max}=40$ dB, at $I_{\mathrm{p}}\gtrsim10$ dB for $P_{\max}=50$
dB, and at $I_{\mathrm{p}}\gtrsim20$ dB for $P_{\max}=60$ dB. Therefore,
the ESSR for both the NOMA and OMA systems becomes constant w.r.t.
$I_{\mathrm{p}}$ (power-constrained regime) at $I_{\mathrm{p}}\gtrsim\left\{ 0,10,20\right\} $
dB for $P_{\max}=\left\{ 40,50,60\right\} $ dB. 

In Fig.~\ref{fig:SumRate_sseVar}, we investigate the effect of channel
estimation error on the ESSR of the NOMA and OMA systems. It can be
noted from the figure that the ESSR for both the NOMA and OMA systems
decreases with an increase in the value of $\sigma_{\epsilon}^{2}$.
This implies that a decrease in the accuracy of channel estimation
has an adverse effect on the system performance. The effect of the
interference-constrained regime is also clearly evident from the figure,
as the value of ESSR at $I_{\mathrm{p}}=0$ dB is the same for $P_{\max}=50$
dB and $P_{\max}=60$ dB. However, for a large value of $I_{\mathrm{p}}$
(10 dB), the benefit of a larger $P_{\max}$ can be noted from the
figure. 

Fig.~\ref{fig:SumRate_PRVar} shows the effect of the number and
distance of PRs on the ESSR of the NOMA system. In this figure, three
different cases are studied. In Case~I, we consider 2 PRs both located
at a distance of 200~m from the ST; in Case~II, we consider 4 PRs
located at $\left\{ 200,205,210,215\right\} $~m from the ST, and
in Case~III, we consider 10 PRs, all located at a distance of 200~m.
Moreover, for all of the cases, we consider two different values of
transmit power: $P_{\max}=40$~dB and 60~dB. It is evident from
the figure that in the interference-constrained regime (i.e., for
small values of $I_{\mathrm{p}}$), an increase in the number of PRs
results in a reduction of the ESSR. However, in this regime, no benefit
is observed from a higher $P_{\max}$ budget. On the other hand, in
the power-constrained regime (i.e., for large values of $I_{p}$),
the ESSR for all the cases becomes the same, irrespective of the number/location
of PRs. This occurs due to the reason that in the power-constrained
regime, the secrecy performance becomes independent of the ST-PR link
quality as well as the interference constraint at the PRs. However,
a larger $P_{\max}$ turns out to be clearly beneficial in this regime. 

\section{Conclusion}

In this paper, we have presented the ergodic secrecy rate analysis
of a two-user downlink NOMA system in an underlay spectrum sharing
scenario consisting of multiple primary-user receivers in the presence
of channel estimation error. Exact and asymptotic closed-form expressions
for the ESSR of the NOMA system were derived. Our results confirmed
that the NOMA system outperforms its OMA-based counterpart in terms
of the ESSR. More interestingly, the results confirmed that no benefit
is obtained in terms of the ESSR from a higher power budget at the
ST in the interference-constrained regime, whereas, in the power-constrained
regime, the ESSR becomes independent of the number of PRs as well
as the quality of the ST-PR links. The results also showed that a
larger channel estimation error results in a reduced secrecy rate.
Finally, the asymptotic analysis demonstrated that the ESSR of the
NOMA system becomes independent of the value of peak tolerable interference
and therefore, the slope of the ESSR w.r.t. the peak tolerable interference
becomes zero. 

\appendices{}

\section{\label{sec:Proof-of-Theorem-RsAvg}Proof of Theorem~\ref{thm:RsAvg}}

Using~\eqref{eq:RsAvgDef}, we have 
\begin{align}
 & \bar{R}_{\mathrm{s}}=\mathbb{E}\left\{ \max\left\{ \log_{2}\left(1+\hat{g}_{\mathrm{s}}P\right)-\log_{2}\left(1+\hat{g}_{\mathrm{e}}P\right),0\right\} \right\} \nonumber \\
= & \int_{y=0}^{\Theta}\int_{x=\sigma_{\epsilon}^{2}}^{\infty}\!\!\!\!\log_{2}\left(1\!+\!P_{\max}x\right)f_{\hat{g}_{\mathrm{s}}}\!\left(x\right)\mathscr{F}_{\hat{g}_{\mathrm{e}}}\!\left(\sigma_{\epsilon}^{2},x\right)f_{\tilde{g}_{\mathrm{p}}}\!\left(y\right)\mathrm{d}x\mathrm{d}y\nonumber \\
+ & \int_{y=\Theta}^{\infty}\int_{x=\sigma_{\epsilon}^{2}}^{\infty}\!\!\!\!\log_{2}\left(1\!+\!I_{\mathrm{p}}\dfrac{x}{y}\right)f_{\hat{g}_{\mathrm{s}}}\!\left(x\right)\mathscr{F}_{\hat{g}_{\mathrm{e}}}\!\left(\sigma_{\epsilon}^{2},x\right)f_{\tilde{g}_{\mathrm{p}}}\!\left(y\right)\mathrm{d}x\mathrm{d}y\nonumber \\
- & \int_{y=0}^{\Theta}\int_{x=\sigma_{\epsilon}^{2}}^{\infty}\!\!\log_{2}\left(1+P_{\max}x\right)f_{\hat{g}_{\mathrm{e}}}\left(x\right)\mathcal{F}_{\hat{g}_{\mathrm{s}}}\left(x\right)f_{\tilde{g}_{\mathrm{p}}}\left(y\right)\mathrm{d}x\mathrm{d}y\nonumber \\
- & \int_{y=\Theta}^{\infty}\int_{x=\sigma_{\epsilon}^{2}}^{\infty}\!\!\log_{2}\left(1+I_{\mathrm{p}}\dfrac{x}{y}\right)f_{\hat{g}_{\mathrm{e}}}\left(x\right)\mathcal{F}_{\hat{g}_{\mathrm{s}}}\left(x\right)f_{\tilde{g}_{\mathrm{p}}}\left(y\right)\mathrm{d}x\mathrm{d}y\nonumber \\
\triangleq & \mathscr{I}_{\mathrm{s,1}}+\mathscr{I}_{s,2}-\mathscr{I}_{s,3}-\mathscr{I}_{s,4}.\label{eq:RsAvgDef-New}
\end{align}
Solving for $\mathscr{I}_{\mathrm{s,1}}$, using the expressions for
$f_{\hat{g}_{\mathrm{s}}}\left(x\right),\mathscr{F}_{\hat{g}_{\mathrm{e}}}\left(\sigma_{\epsilon}^{2},x\right)$
and $f_{\tilde{g}_{\mathrm{p}}}\left(y\right)$ yields 
\begin{align}
 & \mathscr{I}_{\mathrm{s,1}}=-\log_{2}\left(e\right)\sum_{\ell\in\left\{ 1,2,3\right\} }\sum_{\boldsymbol{\eta}\in\Phi}\hat{A}_{\ell}B_{\boldsymbol{\eta}}\kappa_{\boldsymbol{\eta}}\nonumber \\
\times & \left[\int_{y=0}^{\Theta}\int_{x=\sigma_{\epsilon}^{2}}^{\infty}\!\!\ln\left(1\!+\!P_{\max}x\right)\exp\!\left(\!\dfrac{-x}{a_{\mathrm{s}}\Xi_{\ell}}\!\right)\exp\left(-B_{\boldsymbol{\eta}}y\right)\mathrm{d}x\mathrm{d}y\right.\nonumber \\
- & \left.\hat{\mathscr{C}}_{\mathrm{e}}\!\!\int_{y=0}^{\Theta}\!\int_{x=\sigma_{\epsilon}^{2}}^{\infty}\!\!\!\!\!\ln\left(1\!+\!P_{\max}x\right)\exp\!\left(\!\dfrac{-x}{a_{\mathrm{s}}\Xi_{\ell,\mathrm{e}}}\!\right)\!\exp\left(-B_{\boldsymbol{\eta}}y\right)\mathrm{d}x\mathrm{d}y\right]\nonumber \\
= & -\log_{2}\left(e\right)\sum_{\ell\in\left\{ 1,2,3\right\} }\sum_{\boldsymbol{\eta}\in\Phi}\hat{A}_{\ell}\kappa_{\boldsymbol{\eta}}\left\{ 1-\exp\left(-B_{\boldsymbol{\eta}}\Theta\right)\right\} \nonumber \\
 & \qquad\times\left[\int_{x=\sigma_{\epsilon}^{2}}^{\infty}\!\!\ln\left(1\!+\!P_{\max}x\right)\exp\!\left(\!\dfrac{-x}{a_{\mathrm{s}}\Xi_{\ell}}\!\right)\mathrm{d}x\right.\nonumber \\
 & \qquad-\left.\hat{\mathscr{C}}_{\mathrm{e}}\int_{x=\sigma_{\epsilon}^{2}}^{\infty}\!\!\ln\left(1\!+\!P_{\max}x\right)\exp\!\left(\!\dfrac{-x}{a_{\mathrm{s}}\Xi_{\ell,\mathrm{e}}}\!\right)\mathrm{d}x\right].
\end{align}
Computing the integrals above using integration by parts and with
some algebraic manipulations, we arrive at the closed-form expression
for $\mathscr{I}_{\mathrm{s,1}}$ is given by~\eqref{eq:Is1Closed}.
Similarly, for $\mathscr{I}_{\mathrm{s,3}}$, we have 
\begin{align}
 & \mathscr{I}_{\mathrm{s,3}}=-\log_{2}\left(e\right)\hat{C}_{\mathrm{e}}\sum_{\ell\in\left\{ 1,2,3\right\} }\sum_{\boldsymbol{\eta}\in\Phi}\hat{\mathcal{A}}_{\ell}B_{\boldsymbol{\eta}}\kappa_{\boldsymbol{\eta}}\nonumber \\
 & \times\int_{y=0}^{\Theta}\int_{x=\sigma_{\epsilon}^{2}}^{\infty}\!\!\!\!\ln\left(1\!+\!P_{\max}x\right)\exp\left(\!\dfrac{-x}{a_{\mathrm{s}}\Xi_{\ell,\mathrm{e}}}\!\right)\exp\left(-B_{\boldsymbol{\eta}}y\right)\mathrm{d}x\mathrm{d}y\nonumber \\
 & =-\log_{2}\left(e\right)\hat{C}_{\mathrm{e}}\sum_{\ell\in\left\{ 1,2,3\right\} }\sum_{\boldsymbol{\eta}\in\Phi}\hat{\mathcal{A}}_{\ell}\kappa_{\boldsymbol{\eta}}\left\{ 1-\exp\left(-B_{\boldsymbol{\eta}}\Theta\right)\right\} \nonumber \\
 & \qquad\qquad\times\int_{x=\sigma_{\epsilon}^{2}}^{\infty}\!\!\ln\left(1\!+\!P_{\max}x\right)\exp\left(\!\dfrac{-x}{a_{\mathrm{s}}\Xi_{\ell,\mathrm{e}}}\!\right)\mathrm{dx}.\label{eq:Is3Temp}
\end{align}
A closed-form expression for \eqref{eq:Is3Temp} can be obtained by
following the steps similar to those used to obtain $\mathscr{I}_{\mathrm{s,1}}$,
and is given in~\eqref{eq:Is3Closed}. Now solving for $\mathscr{I}_{\mathrm{s,4}}$,
we have 
\begin{align}
 & \mathscr{I}_{s,4}=-\log_{2}\left(e\right)\hat{C}_{\mathrm{e}}\sum_{\ell\in\left\{ 1,2,3\right\} }\sum_{\boldsymbol{\eta}\in\Phi}\hat{\mathcal{A}}_{\ell}B_{\boldsymbol{\eta}}\kappa_{\boldsymbol{\eta}}\int_{y=\Theta}^{\infty}\nonumber \\
 & \left[\int_{x=\sigma_{\epsilon}^{2}}^{\infty}\ln\left(1+\dfrac{I_{\mathrm{p}}}{y}x\right)\exp\left(\dfrac{-x}{a_{\mathrm{s}}\Xi_{\ell,\mathrm{e}}}\right)\mathrm{d}x\right]\exp\left(-B_{\boldsymbol{\eta}}y\right)\mathrm{d}y.\label{eq:Is4Temp1}
\end{align}
Integrating the preceding expression w.r.t. $x$ using integration
by parts and using a few algebraic manipulations yields 
\begin{align}
 & \mathscr{I}_{\mathrm{s,4}}=-\log_{2}\left(e\right)\hat{C}_{\mathrm{e}}\sum_{\ell\in\left\{ 1,2,3\right\} }\sum_{\boldsymbol{\eta}\in\Phi}\hat{\mathcal{A}}_{\ell}B_{\boldsymbol{\eta}}\kappa_{\boldsymbol{\eta}}a_{\mathrm{s}}\Xi_{\ell,\mathrm{e}}\int_{y=\Theta}^{\infty}\nonumber \\
 & \times\left[\exp\left(\dfrac{-\sigma_{\epsilon}^{2}}{a_{\mathrm{s}}\Xi_{\ell,\mathrm{e}}}\right)\ln\left(1+\dfrac{I_{\mathrm{p}}\sigma_{\epsilon}^{2}}{y}\right)+\exp\left(\dfrac{y}{a_{\mathrm{s}}I_{\mathrm{p}}\Xi_{\ell,\mathrm{e}}}\right)\right.\nonumber \\
 & \left.\times\mathrm{E}_{1}\left(\dfrac{1}{a_{\mathrm{s}}\Xi_{\ell,\mathrm{e}}}\left(\dfrac{y}{I_{\mathrm{p}}}+\sigma_{\epsilon}^{2}\right)\right)\right]\exp\left(-B_{\boldsymbol{\eta}}y\right)\mathrm{d}y.\label{eq:Is4Temp2}
\end{align}
Now solving the integral in~\eqref{eq:Is4Temp2} w.r.t $y$ using
integration by parts, a closed-form expression for $\mathscr{I}_{\mathrm{s,4}}$
is given in~\eqref{eq:Is4Closed}. Next, for $\mathscr{I}_{\mathrm{s,2}},$
we have 
\begin{align}
 & \mathscr{I}_{\mathrm{s,}2}\!=\!-\log_{2}\left(e\right)\!\!\!\sum_{\ell\in\left\{ 1,2,3\right\} }\!\sum_{\boldsymbol{\eta}\in\Phi}\!\hat{A}_{\ell}B_{\boldsymbol{\eta}}\kappa_{\boldsymbol{\eta}}\int_{y=\Theta}^{\infty}\int_{x=\sigma_{\epsilon}^{2}}^{\infty}\!\!\!\ln\left(\!1+\dfrac{I_{\mathrm{p}}}{y}x\!\right)\nonumber \\
 & \times\exp\left(\dfrac{-x}{a_{\mathrm{s}}\Xi_{\ell}}\right)\left\{ 1-\hat{\mathscr{C}}_{\mathrm{e}}\exp\left(\dfrac{-x}{a_{\mathrm{s}}\tilde{\Omega}_{\mathrm{e}}}\right)\right\} \exp\left(-B_{\boldsymbol{\eta}}y\right)\mathrm{d}x\mathrm{d}y.\label{eq:Is2Temp}
\end{align}
Solving the integrals above by following the steps similar to those
used for $\mathscr{I}_{\mathrm{s,4}}$, a closed-form expression for
$\mathscr{I}_{\mathrm{s,2}}$ is given by~\eqref{eq:Is2Closed}.
This concludes the proof. 

\section{\label{sec:Proof-of-Theorem-RwAvg}Proof of Theorem~\ref{thm:RwAvg}}

Using~\eqref{eq:RwAvgDef}, we have 
\begin{align}
 & \bar{R}_{\mathrm{w}}=\mathbb{E}\left\{ \max\left\{ \log_{2}\left(\dfrac{1+\left(\tilde{g}_{_{\mathrm{w}}}+\sigma_{\epsilon}^{2}\right)P}{1+\left(a_{\mathrm{s}}\tilde{g}_{\mathrm{_{\mathrm{w}}}}+\sigma_{\epsilon}^{2}\right)P}\right)\right.\right.\nonumber \\
 & \qquad\qquad\qquad-\left.\left.\log_{2}\left(\dfrac{1+\left(\tilde{g}_{_{\mathrm{e}}}+\sigma_{\epsilon}^{2}\right)P}{1+\left(a_{\mathrm{s}}\tilde{g}_{_{\mathrm{e}}}+\sigma_{\epsilon}^{2}\right)P}\right),0\right\} \right\} \nonumber \\
 & =\left[\int_{y=0}^{\Theta}\int_{x=\sigma_{\epsilon}^{2}}^{\infty}\!\!\!\!\!\log_{2}\left(1\!+\!xP_{\max}\right)f_{\check{g}_{_{\mathrm{w}}}}\!\left(x\right)\mathscr{F}_{\check{g}_{_{\mathrm{e}}}}\!\left(x\right)f_{\tilde{g}_{_{\mathrm{p}}}}\!\left(y\right)\mathrm{d}x\mathrm{d}y\right.\nonumber \\
 & \left.-\int_{y=0}^{\Theta}\int_{x=\sigma_{\epsilon}^{2}}^{\infty}\!\!\!\!\!\log_{2}\left(1\!+\!xP_{\max}\right)f_{\hat{g}_{_{\mathrm{w}}}}\!\left(x\right)\mathscr{F}_{\hat{g}_{_{\mathrm{e}}}}\!\left(x\right)f_{\tilde{g}_{_{\mathrm{p}}}}\!\left(y\right)\mathrm{d}x\mathrm{d}y\right]\nonumber \\
 & +\left[\int_{y=\Theta}^{\infty}\int_{x=\sigma_{\epsilon}^{2}}^{\infty}\!\!\!\!\!\log_{2}\left(1+I_{\mathrm{p}}\dfrac{x}{y}\right)f_{\check{g}_{_{\mathrm{w}}}}\!\left(x\right)\mathscr{F}_{\check{g}_{_{\mathrm{e}}}}\!\left(x\right)f_{\tilde{g}_{_{\mathrm{p}}}}\!\left(y\right)\mathrm{d}x\mathrm{d}y\right.\nonumber \\
 & \left.-\int_{y=\Theta}^{\infty}\int_{x=\sigma_{\epsilon}^{2}}^{\infty}\!\!\!\!\!\log_{2}\left(1+I_{\mathrm{p}}\dfrac{x}{y}\right)f_{\hat{g}_{_{\mathrm{w}}}}\!\left(x\right)\mathscr{F}_{\hat{g}_{_{\mathrm{e}}}}\!\left(x\right)f_{\tilde{g}_{_{\mathrm{p}}}}\!\left(y\right)\mathrm{d}x\mathrm{d}y\right]\nonumber \\
 & -\left[\int_{y=0}^{\Theta}\int_{x=\sigma_{\epsilon}^{2}}^{\infty}\!\!\!\!\!\log_{2}\left(1\!+\!xP_{\max}\right)f_{\check{g}_{_{\mathrm{e}}}}\!\left(x\right)\mathcal{F}_{\check{g}_{_{\mathrm{w}}}}\!\left(x\right)f_{\tilde{g}_{_{\mathrm{p}}}}\!\left(y\right)\mathrm{d}x\mathrm{d}y\right.\nonumber \\
 & \left.-\int_{y=0}^{\Theta}\int_{x=\sigma_{\epsilon}^{2}}^{\infty}\!\!\!\!\!\log_{2}\left(1\!+\!xP_{\max}\right)f_{\hat{g}_{_{\mathrm{e}}}}\!\left(x\right)\mathcal{F}_{\hat{g}_{_{\mathrm{w}}}}\!\left(x\right)f_{\tilde{g}_{_{\mathrm{p}}}}\!\left(y\right)\mathrm{d}x\mathrm{d}y\right]\nonumber \\
 & -\left[\int_{y=\Theta}^{\infty}\int_{x=\sigma_{\epsilon}^{2}}^{\infty}\!\!\!\!\!\log_{2}\left(1+I_{\mathrm{p}}\dfrac{x}{y}\right)f_{\check{g}_{_{\mathrm{e}}}}\!\left(x\right)\mathcal{F}_{\check{g}_{_{\mathrm{w}}}}\!\left(x\right)f_{\tilde{g}_{_{\mathrm{p}}}}\!\left(y\right)\mathrm{d}x\mathrm{d}y\right.\nonumber \\
 & \left.-\int_{y=\Theta}^{\infty}\int_{x=\sigma_{\epsilon}^{2}}^{\infty}\!\!\!\!\!\log_{2}\left(1+I_{\mathrm{p}}\dfrac{x}{y}\right)f_{\hat{g}_{_{\mathrm{e}}}}\!\left(x\right)\mathcal{F}_{\hat{g}_{_{\mathrm{w}}}}\!\left(x\right)f_{\tilde{g}_{_{\mathrm{p}}}}\!\left(y\right)\mathrm{d}x\mathrm{d}y\right]\nonumber \\
 & \triangleq\mathscr{I}_{\mathrm{w,1}}+\mathscr{I}_{\mathrm{w,2}}-\mathscr{I}_{\mathrm{w,3}}-\mathscr{I}_{\mathrm{w,4}}.\label{eq:RwAvgDef-New}
\end{align}
Following a similar line of argument as given in~Appendix~\ref{sec:Proof-of-Theorem-RsAvg},
closed-form expressions for $\mathscr{I}_{\mathrm{w,1}},\mathscr{I}_{\mathrm{w,2}},\mathscr{I}_{\mathrm{w,3}}$
and $\mathscr{I}_{\mathrm{w,4}}$ are given by~\eqref{eq:Iw1Closed}
\textendash{} \eqref{eq:Iw4Closed}. This completes the proof. 

\IEEEtriggeratref{5}

\bibliographystyle{IEEEtran}
\bibliography{Invited}

\end{document}